\documentclass[twocolumn,prl,showpacs,preprintnumbers,amsmath,amssymb]{revtex4} %prl,twocolumn
\usepackage{graphicx}% Include figure files
\usepackage{dcolumn}% Align table columns on decimal point
\usepackage{bm}% bold math
\begin{document}
\title{A Generalized Ramsey Excitation Scheme with Suppressed Light Shift}
\author{N. Huntemann}
\email{nils.huntemann@ptb.de}
\author{B. Lipphardt}
\author{M. Okhapkin}
\author{Chr. Tamm}
\author{E. Peik}
\affiliation{Physikalisch-Technische Bundesanstalt, Bundesallee 100, 38116 Braunschweig, Germany}

\author{A. V. Taichenachev and V. I. Yudin}
\affiliation{Institute of Laser Physics SB RAS, Novosibirsk 630090, Russia} \affiliation{Novosibirsk State University, Novosibirsk 630090, Russia} 
\affiliation{Novosibirsk State Technical University, Novosibirsk 630092, Russia}
\date{\today}
\begin{abstract}
We experimentally investigate a recently proposed optical excitation scheme [V.I. Yudin \textit{et al.}, Phys. Rev. A 82, 011804(R)(2010)] that is a generalization of Ramsey's method of separated oscillatory fields and consists of a sequence of three excitation pulses. The pulse sequence is tailored to produce a resonance signal that is immune to the light shift and other shifts of the transition frequency that are correlated with the interaction with the probe field. We investigate the scheme using a single trapped $^{171}$Yb$^+$ ion and excite the highly forbidden $^2S_{1/2}-^2F_{7/2}$ electric-octupole transition under conditions where the light shift is much larger than the excitation linewidth, which is in the hertz range. The experiments demonstrate a suppression of the light shift by four orders of magnitude and an immunity against its fluctuations. 
\end{abstract}

%\keywords{}

\maketitle

Ramsey's method of separated oscillatory fields was crucial for the progress in precision spectroscopy and the development of atomic clocks \cite{Ramsey1990} and is an important tool in quantum information processing \cite{Haeffner2008}. In the Ramsey scheme, two levels of a quantum system are brought into a coherent superposition by a first excitation pulse followed by a free evolution period. After a second excitation pulse the population in one of the levels is detected, which shows the effect of the interference of the second pulse with the time-evolved superposition state. In the original experiments with atomic and molecular beams, this permitted the recording of a resonance line shape with a width that is mainly determined by the total interaction time, without shifts and broadening through inhomogeneous excitation conditions. Ramsey's method is employed mainly to excite states with a natural lifetime exceeding the interaction time, so that primarily transitions forbidden by electric dipole selection rules are investigated. Especially in the case of optical spectroscopy the high probe light intensities required to drive these transitions will unavoidably lead to level shifts through the dynamical Stark effect. This so-called light shift appears through the nonresonant coupling to other energy levels by the probe light and is usually proportional to its intensity. Several methods were investigated to compensate this shift, for example, linear extrapolation to zero intensity or the use of an additional inversely shifting field \cite{Haeffner2003}. Nevertheless, a wide range of precise frequency measurements presently suffer from significant uncertainties due to light shift. Here, two-photon \cite{Millerioux1994,Badr2006,Zanon2006,Parthey2011}, higher order multipole \cite{King2012,Huntemann2012}, and magnetic-field induced transitions \cite{Taichenachev2006,Baillard2007,Poli2008,Akatsuka2010} are good examples. 

\begin{figure}
\includegraphics[width=\columnwidth]{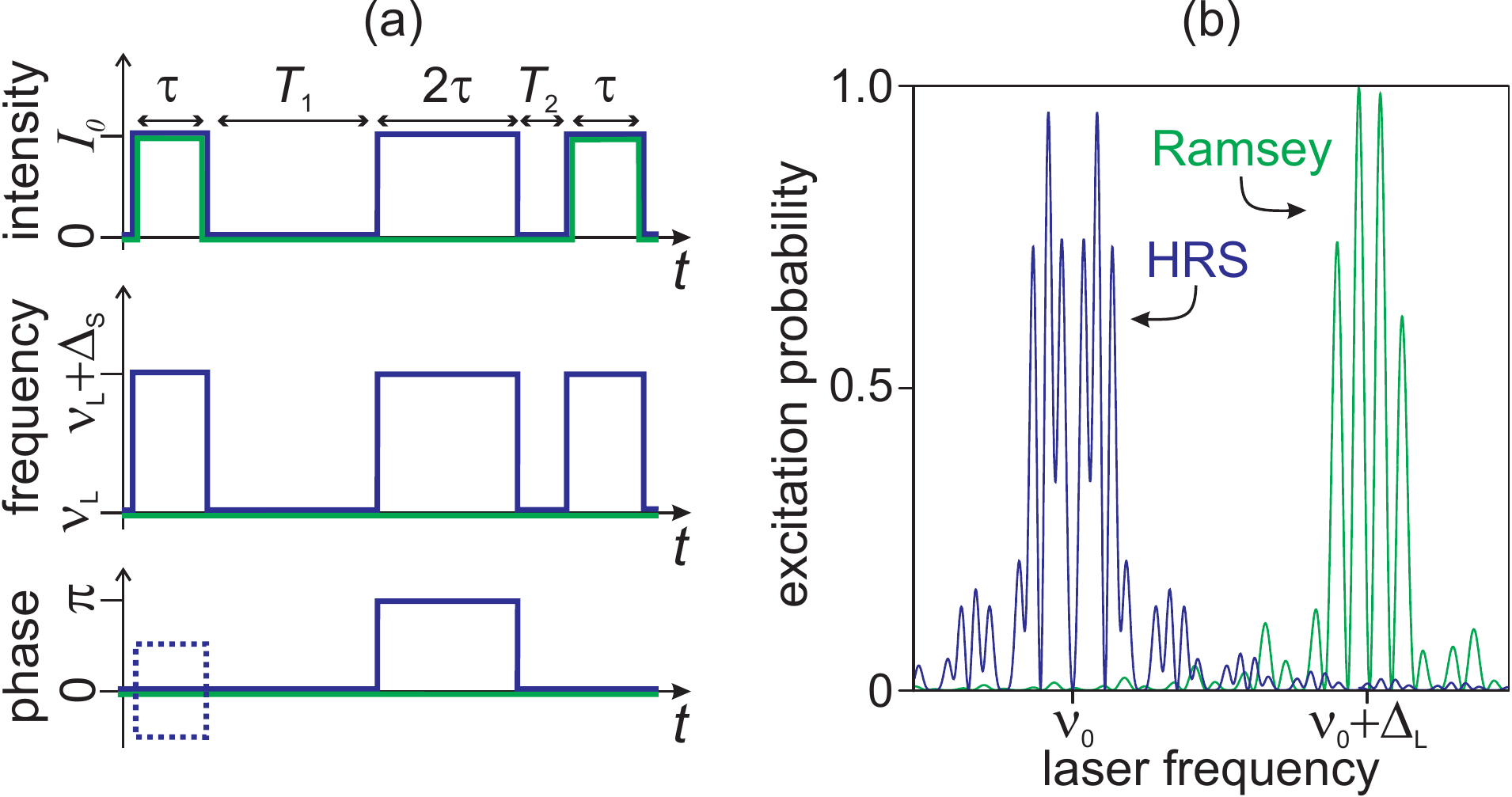}
\caption{Pulse sequence (a) and resulting excitation spectrum (b) of the Ramsey (green) and the ``hyper-Ramsey'' spectroscopy (HRS) excitation scheme (blue). Here $\nu_\text{L}$ is the probe laser frequency and $\nu_0$ the unperturbed transition frequency. The laser step frequency $\Delta_\text{S}$ is assumed to be equal to the light shift $\Delta_\text{L}$ and the intensity $I_0$ is chosen to obtain a pulse area $\pi/2$ for a pulse duration $\tau$. A discriminator signal can be generated by alternately stepping the phase of the first pulse by $\pm\pi/2$ as indicated by the dotted lines. The spectra are calculated for the parameters $T_1=2\tau$, $T_2=0$, $\Delta_\text{L}=4.1/\tau$ with equal dark period durations in both schemes. \label{Schematic}}
\end{figure}  

In this Letter we present the first experimental realization of a generalized Ramsey excitation scheme, the so-called ''hyper-Ramsey'' spectroscopy (HRS) recently proposed by Yudin \textit{et al.} \cite{Yudin2010}, which cancels the light shift and efficiently suppresses the sensitivity of the spectroscopic signal to variations of the probe light intensity.
This scheme and a typical resulting excitation spectrum are compared to conventional Ramsey spectroscopy in Fig.~\ref{Schematic}. 
The spectrum obtained with Ramsey excitation usually shows indications of the presence of light shift: the position and shape of the envelope reflects the excitation spectrum resulting from one of the pulses, whereas the Ramsey fringes result from coherent excitation with both pulses and the intermediate dark period. The fringes are less shifted than the envelope, because their shift is determined by the time average of the intensity. This results in a shifted and asymmetric Ramsey pattern [see Fig. \ref{Schematic}(b)], as observed in early demonstrations of optical Ramsey spectroscopy of a two-photon transition \cite{Hollberg1984}. 

This intuitive picture suggests that the effect of the light shift $\Delta_\text{L}$ on the spectrum can be compensated by introducing a frequency step of the probe light $\Delta_\text{S}=\Delta_\text{L}$ during the interrogation pulses whereas the unshifted probe frequency $\nu_\text{L}$ is tuned across the resonance with the unperturbed atomic frequency $\nu_0$, as proposed in \cite{Taichenachev2009}. The scheme can be made additionally insensitive against small changes of the laser intensity or errors in $\Delta_\text{S}$ by inserting an additional pulse with identical intensity and frequency and with a doubled duration between the Ramsey pulses. This feature bears resemblance to ``echo'' techniques \cite{Warren1983} that can be used to suppress dephasing between the atomic oscillators during the free evolution. Here, however, the additional pulse compensates the dephasing between the atomic coherence and the probe laser field caused by the Ramsey pulses. The phase of the additional pulse is shifted by $\pi$ relative to the Ramsey pulses in order to improve the robustness against variations of the pulse area. The frequency steps are applied in a phase-coherent way so that they do not introduce additional phase changes of the probe field. These are the essential elements of the HRS scheme \cite{Yudin2010}. The corresponding probe pulse pattern is sketched in Fig.~\ref{Schematic}(a).  

According to Ref.~\cite{Yudin2010}, the line shape of the HRS resonance signal can be calculated as the excitation probability $p=|\langle e|\widehat{M}|g\rangle|^2$ from the atomic ground state $|g\rangle= \tbinom{0}{1}$ to the excited state $|e\rangle= \tbinom{1}{0}$. Here $\widehat{M}$ describes the effect of the sequence of excitation pulses $\widehat{W}(\tau,\Omega_0,\Delta\nu_0+\Delta_\text{L}-\Delta_\text{S})$ and dark periods $\widehat{V}(T,\Delta\nu_0)$: 
\begin{equation}
\begin{split}
\widehat{M}&= \widehat{W}(\tau,\Omega_0,\Delta\nu_0+\Delta_\text{L}-\Delta_\text{S}) \cdot \widehat{V}(T_2,\Delta\nu_0)  \\ 
&\quad\cdot \widehat{W}(2\tau,-\Omega_0,\Delta\nu_0+\Delta_\text{L}-\Delta_\text{S}) \cdot \widehat{V}(T_1,\Delta\nu_0) \\
&\quad\cdot \widehat{W}(\tau,e^{i\phi}\Omega_0,\Delta\nu_0+\Delta_\text{L}-\Delta_\text{S}).
\label{HRSscheme}
\end{split}
\end{equation}
Here $\Delta\nu_0$ denotes the detuning of the probe laser from the unperturbed transition frequency $\nu_0$. The relative optical phase during each pulse is indicated by phase factors of the resonant Rabi frequency $\Omega_0$. In Fig.~\ref{Schematic}(a) and Eq.~\ref{HRSscheme} it is assumed that the total dark period $T$ is divided into two parts, $T=T_1+T_2$. Calculations show that the contrast of the HRS resonance signal and the light shift suppression is maximal for either $T_1=0$ or  $T_2=0$. Introducing two dark periods may be technically advantageous because in this case both phase reversals can be carried out when the probe light is switched off, which excludes the possibility that the phase reversal leads to transient pulse distortions. For simplicity, the case $T_2=0$ will be assumed in the following. 

\begin{figure}
\includegraphics[width=\columnwidth]{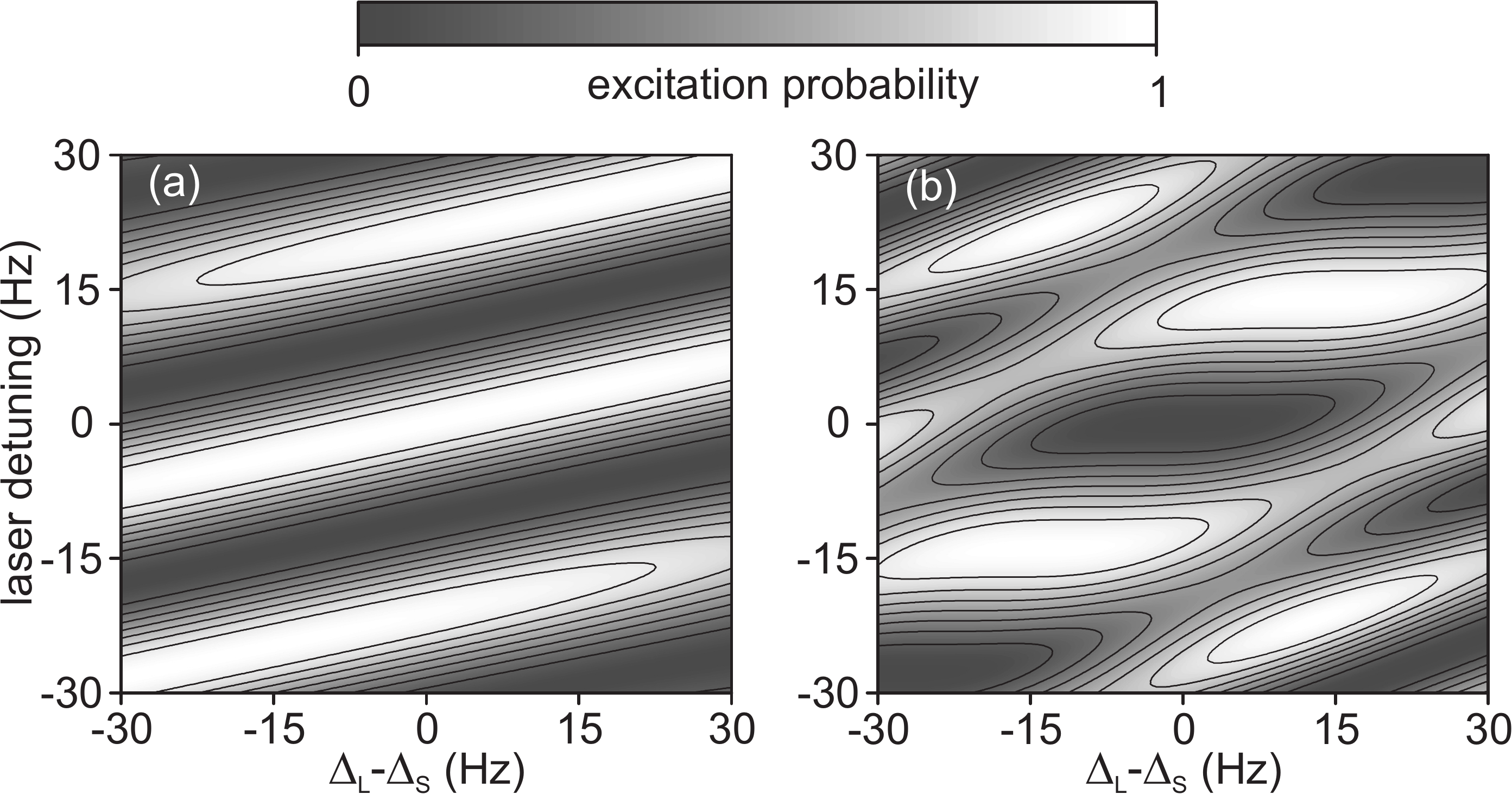}
\caption{Calculated excitation probability as a function of the uncompensated light shift $\Delta_\text{L}-\Delta_\text{S}$ and of the laser detuning $\Delta\nu_0$ from the unperturbed transition frequency for (a) Ramsey and (b) HRS excitation. In accordance with typical parameters of the performed experiment, the duration of the $\pi/2$ pulse is $\tau=9$~ms and that of the dark period $T=36$~ms. 
\label{RamseyvsHyper}}
\end{figure}

Typical spectra calculated for the Ramsey and the HRS schemes are shown in Fig.~\ref{Schematic}(b). The remarkable result of the theoretical analysis \cite{Yudin2010} is that the linear dependence of the frequency of the central minimum of the HRS resonance signal on an error in the compensation frequency $\Delta_\text{S}$ can be eliminated over a range proportional to $\Omega_0$. The immunity against an uncompensated light shift $\Delta_\text{L}-\Delta_\text{S}$ is illustrated in Fig.~\ref{RamseyvsHyper} by comparing the spectral dependence of the excitation probability for the Ramsey and HRS pulse sequences. For this comparison, a frequency step $\Delta_\text{S}\neq0$ during the pulses was also used for the Ramsey case as in Ref.~\cite{Taichenachev2009}. Here the fringe pattern position depends linearly on the residual light shift $\Delta_\text{L}-\Delta_\text{S}$ with a slope of $(1+\pi T/4\tau)^{-1}$. In contrast to this, the HRS  excitation shows a distinctly nonlinear dependence, so that the frequency of the minimum excitation probability remains nearly constant in a wide interval around the perfect compensation condition $\Delta_\text{S}=\Delta_\text{L}$. Beyond that range the position of the HRS fringe pattern shifts with a slope that is larger than in the case of Ramsey excitation. The spectral resolution obtained with the HRS scheme is only slightly reduced compared to Ramsey excitation. 

In an atomic frequency standard the frequency of an oscillator is stabilized to the line center of the atomic resonance signal. In most cases this is achieved by alternately recording resonance signals with a fixed positive and negative detuning around the line center. The difference of the excitation probabilities obtained with both detunings yields a discriminator signal that varies antisymmetrically around the resonance center and can be used for stabilization. For Ramsey excitation, a discriminator signal can also be produced by alternately applying phase steps of $\phi=\pm\pi/2$ to one of the excitation pulses while the excitation frequency is kept constant \cite{Ramsey1951,Letchumanan2004}. If applied to the HRS excitation scheme as shown in Fig.~\ref{Schematic}(a), the latter technique is particularly advantageous because the immunity to light shift fluctuations is further enhanced (see below). 

To experimentally investigate the HRS scheme we apply it to excite the highly forbidden electric-octupole transition $^2S_{1/2}-{^2F_{7/2}}$ in $^{171}\text{Yb}^+$ at a wavelength of 467~nm. Here, the light shift induced by the probe laser field that excites the transition is typically much larger than the resolved linewidth. This frequency shift leads to one of the dominant contributions to the uncertainty of an optical clock based on the octupole transition \cite{Huntemann2012,King2012}, and therefore application of the HRS scheme promises a significant improvement.

Our experimental setup is described in detail in Refs.~\cite{Tamm2009,Huntemann2012}. A laser-cooled single $^{171}\text{Yb}^+$ ion is confined in a Paul trap. Each measurement cycle starts with a cooling period (10~ms) followed by hyperfine pumping to the $^2S_{1/2} (F=0)$ ground state (25~ms). Subsequently, the cooling and repumping lasers are blocked and the HRS probe pulse sequence is applied for excitation to the $^2F_{7/2} (F=3,m_F=0)$ state. Absence of the fluorescence at the beginning of the following cooling period indicates population of this state. The excitation probability is determined from multiple repetitions of the sequence. The duration of the probe pulse sequence can be extended up to $\approx 300$~ms, limited by the coherence time of the probe laser, which is in turn limited by thermomechanical noise of the reference cavity to which the laser frequency is stabilized. The coherence time was verified by Rabi excitation using a single 335~ms pulse that yields a Fourier-transform limited linewidth of $2.4$~Hz and a resonant excitation probability of more than 90\%. Up to 10~mW of probe laser power are focused to a beam waist diameter of $40~\mu$m at trap center, permitting a minimum $\pi$-pulse excitation time of $18$~ms. Fluctuations of the probe field intensity at the position of the ion are smaller than 1\% over several hours. The HRS pulse sequence is shaped by an acousto-optic modulator (AOM) driven at a frequency of 80~MHz by a direct digital synthesizer, which enables fast and precise control of the intensity, frequency, and phase of the probe light field. The AOM drive power is switched off only briefly to create precisely defined pulse edges, and a mechanical shutter blocks the probe laser light during the longest parts of the dark periods. In this way, spurious variations of the optical phase induced by temperature changes of the AOM crystal over a probe cycle can be avoided. 

The frequency of the probe laser is constant within 1~Hz over intervals $\leq 100$~s and typically exhibits less than 5~Hz variation after several hours. This frequency stability is achieved by compensating a linear frequency drift of $\approx 30$~mHz/s and by a precise control of the temperature of the reference cavity, which is set close to the value where its coefficient of thermal expansion crosses zero. The laser frequency was tracked with a fiber-laser-based frequency comb generator \cite{Lipphardt2009} using the caesium fountain clock CSF1 of PTB as the reference \cite{Weyers2001}. The light shift $\Delta_\text{L}$ is determined by stabilizing the laser frequency to the resonance signal obtained with Rabi excitation and comparing it to the unperturbed transition frequency \cite{Huntemann2012}.

\begin{figure}
\includegraphics[width=.90\columnwidth]{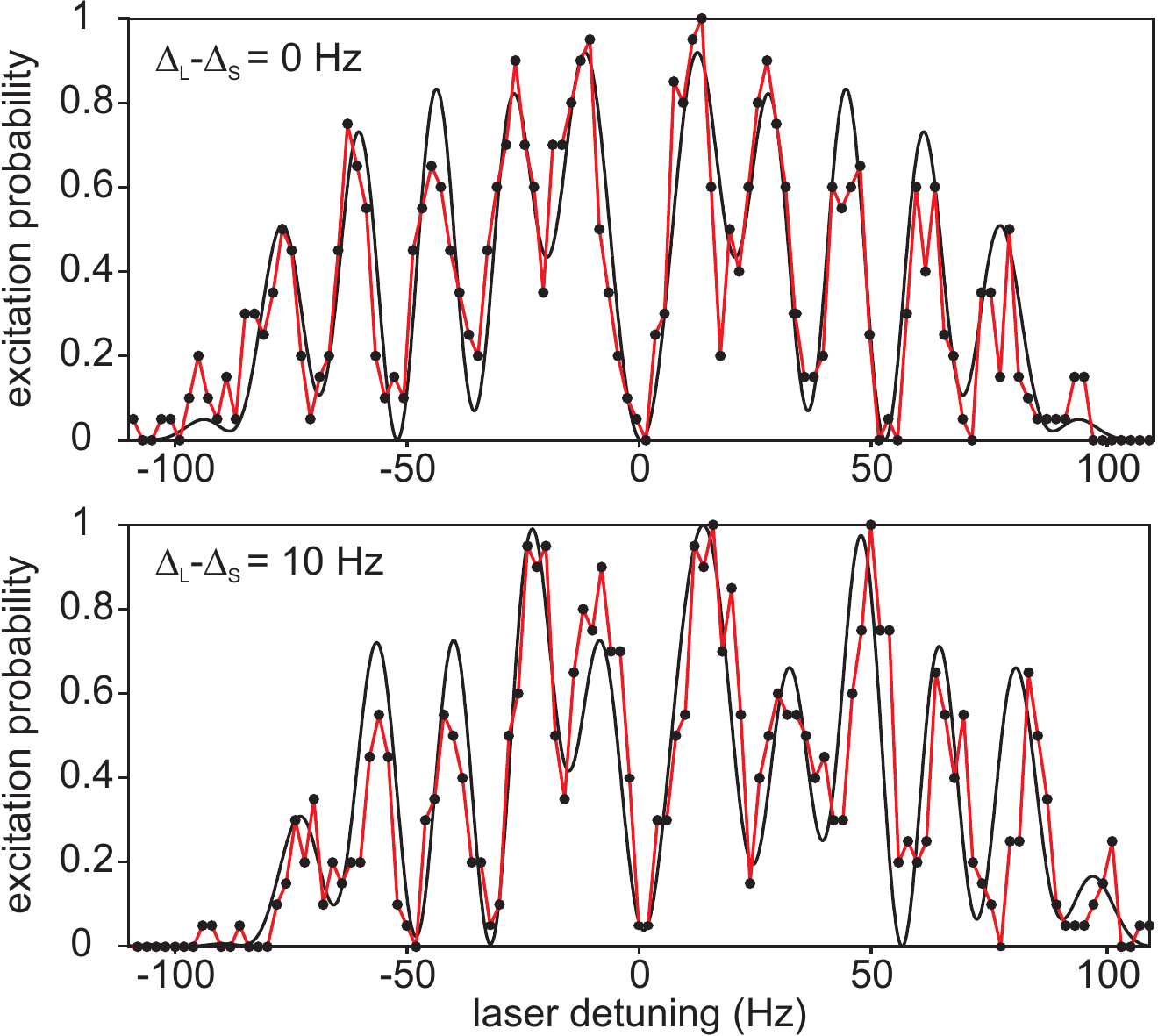}
\caption{Excitation spectra of the octupole transition in $^{171}\text{Yb}^+$ using the HRS scheme with $\tau=9$~ms and $T=36$~ms with fully $(\Delta_\text{L}=\Delta_\text{S})$ and partially $(\Delta_\text{L}\neq\Delta_\text{S})$ compensated light shift for $\Delta_\text{L}\approx1$~kHz. The laser detuning is relative to the unperturbed transition frequency and the data points are the result of 20 interrogations at each frequency step. The solid black lines show the calculated line shapes.
\label{ExcitationSpectra}}
\end{figure}

Figure 3 shows experimental spectra obtained with the HRS scheme together with the calculated fringe patterns for full light shift compensation $(\Delta_\text{L}=\Delta_\text{S})$ and for the case $\Delta_\text{L}-\Delta_\text{S}=10$~Hz, with a light shift $\Delta_\text{L}=1090$~Hz in both cases. From a comparison of recorded spectra with calculated line shapes, $\Delta_\text{L}-\Delta_\text{S}$ was determined with an uncertainty of 1~Hz. The scattering of the measured excitation probabilities is predominantly determined by quantum projection noise. Comparing the two cases, large differences appear in the shape of the excitation spectrum except for the position of the central minimum that is largely unaffected. 

\begin{figure}
\includegraphics[width=.90\columnwidth]{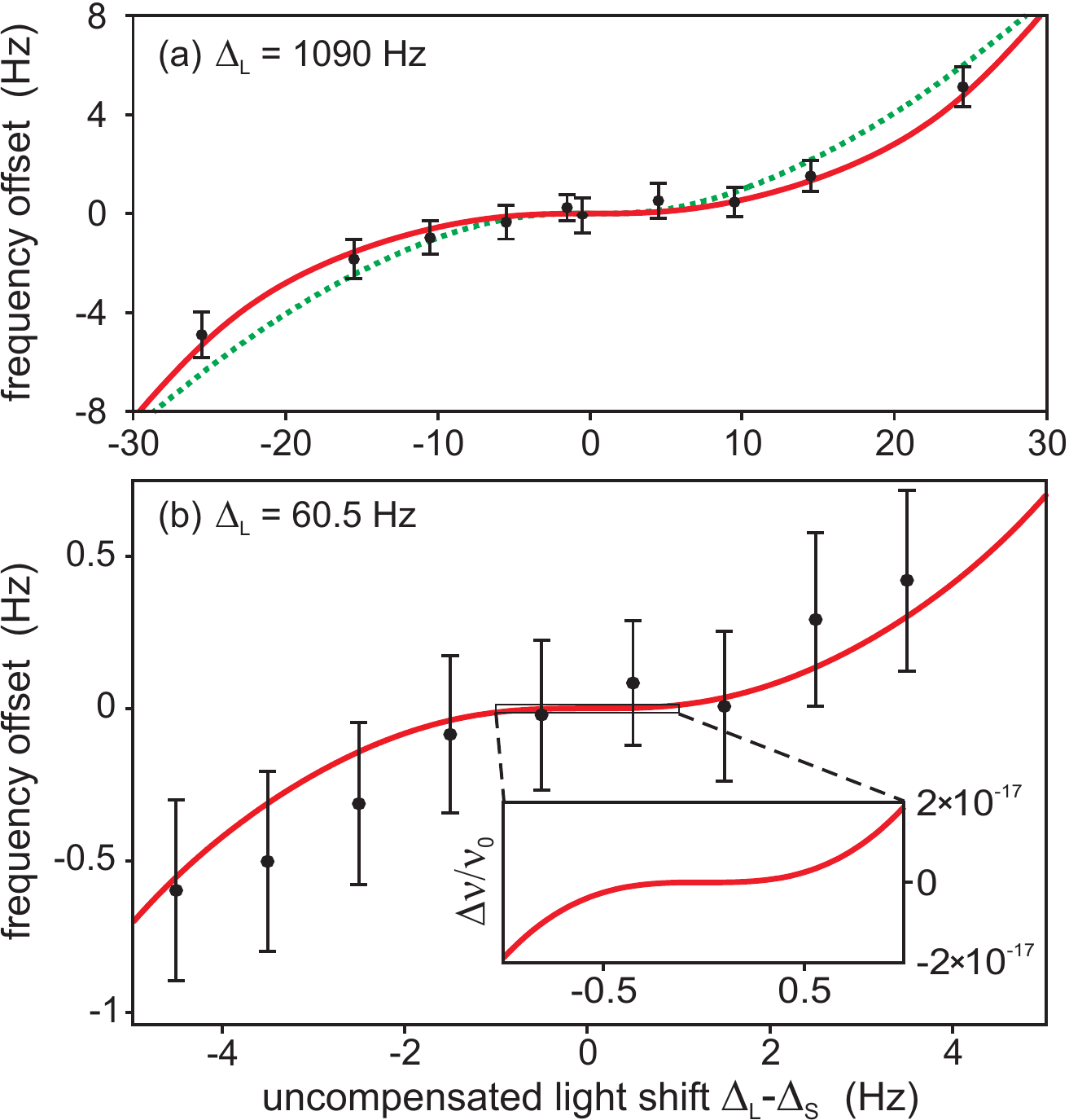}
\caption{Frequency offset of the probe laser stabilized at $\Delta_\text{L}-\Delta_\text{S}$ relative to the fully compensated case $\Delta_\text{S}=\Delta_\text{L}$, for conditions (a) $T=36$~ms, $\tau=9$~ms, and (b) $T=144$~ms, $\tau=36$~ms, corresponding to a linewidth of the central HRS fringe of $3.2$~Hz. The solid red line indicates the predicted dependence if the discriminator signal of the stabilization is generated by alternately stepping the phase of the initial pulse by $\pm\pi/2$. The dashed line in (a) shows the position of the central minimum of the HRS spectrum. The inset in (b) is an enlarged view showing the frequency offset in units of the  frequency $\nu_0$ of the Yb$^+$ octupole transition.
\label{CubicDep}}
\end{figure}

In the following we investigate the predicted nonlinear dependence of the stabilized probe laser frequency on the uncompensated light shift $\Delta_\text{L}-\Delta_\text{S}$.  The discriminator signal for stabilization to the atomic resonance is generated by alternately stepping the phase of the first probe light pulse by $\pm\pi/2$ as described above. We use an interleaved servo technique where a set of parameters is alternated between two settings every fourth measurement cycle and the probe light frequency is stabilized by the respective discriminator signals in two independent digital servo loops \cite{Tamm2009}. This allows us to measure the frequency difference with an essentially quantum-projection noise limited uncertainty of about $5\times 10^{-15}\nu_0 (t/\text{s})^{-1/2}$, where $t$ is the averaging time. The measurements presented in Fig.~\ref{CubicDep} show the frequency offset resulting from variation of $\Delta_\text{S}$ relative to the case of complete light shift compensation, $\Delta_\text{L}=\Delta_\text{S}$. Two different settings were realized: in Fig.~\ref{CubicDep}(a), the duration of the probe pulse is reduced to the limit given by the maximal available probe light intensity and in Fig.~\ref{CubicDep}(b), the probe pulse duration is extended to near the coherence time of the probe laser. In both cases the experimental results are in very good agreement with the calculated dependence that is predominantly cubic around $\Delta_\text{S}=\Delta_\text{L}$. Since the light shift $\Delta_\text{L}$ scales quadratically with the Rabi frequency $\Omega_0$, and the range of efficient suppression is limited to $\left|\Delta_\text{L}-\Delta_\text{S}\right|<\Omega_0$, the relative range $\left|\Delta_\text{L}-\Delta_\text{S}\right|/\Delta_\text{L}$ increases if $\Omega_0$ is reduced. Also shown in Fig.~\ref{CubicDep}(a) is the predicted dependence of the central minimum of the HRS spectrum on the residual light shift. The comparison of the two calculated curves with the experimental data confirms the superior light shift suppression that is obtained if the laser frequency is stabilized by a discriminator signal produced by phase steps of $\phi=\pm\pi/2$ of the first excitation pulse \cite{Yudin2010}. 

The inset of Fig.~\ref{CubicDep}(b) indicates the expected residual relative variation of the stabilized laser frequency if $|\Delta_\text{L}-\Delta_\text{S}|$ is kept within a range of $\pm 1$~Hz. In order to fulfill this condition, a stabilization using the HRS scheme can be combined with a second servo system where Rabi spectroscopy with the same probe light intensity is used and the frequency offset from the HRS stabilization is determined. This offset can be used as an estimate of $\Delta_\text{L}$ in the HRS excitation so that the combined action of the two servo systems minimizes $\left|\Delta_\text{L}-\Delta_\text{S}\right|$ and ensures that slow variations of $\Delta_\text{L}$ will not degrade the light shift suppression. With application of this method for the conditions of Fig.~\ref{CubicDep}(b), the remaining shift of the Yb$^+$ octupole transition frequency by the probe laser field is expected to be well below $10^{-17}$, since $\left|\Delta_\text{L}-\Delta_\text{S}\right|$ can easily be reduced to less than 0.5~Hz using Rabi excitation with a pulse duration of $2\tau=72$~ms. This constitutes a shift suppression by four orders of magnitude. 

In conclusion, we have demonstrated the efficient elimination of the probe laser induced light shift and of its fluctuations using a generalized Ramsey excitation scheme. In contrast to extrapolation techniques this scheme does neither require precise intensity measurements nor information on the relation between intensity and shift. We note that an analogous suppression is expected for any transition frequency shift that appears synchronously with the interaction with the probe light. One well-known example for this is the Zeeman shift in optical frequency standards relying on magnetic-field induced transitions \cite{Taichenachev2006,Baillard2007,Poli2008,Akatsuka2010}. We expect that the application of the HRS method can be advantageous in the precision laser spectroscopy of atoms and molecules and also in quantum information processing if coherence times are limited by light shift fluctuations. 
  
We thank U. Sterr for helpful discussions and S. Weyers for providing the reference for frequency measurements. This work was supported by DFG through QUEST, DFG/RFBR (Grant No. 10-02-91335). A.V.T. and V.I.Y. were supported by RFBR (No. 10-02-00406, No. 11-02-00775, No. 11-02-01240, No. 12-02-00454), programs of RAS, and the federal program ``Scientific and pedagogic personnel of innovative Russia 2009-–2013.''


\begin{thebibliography}{99}

\bibitem{Ramsey1990} N. F. Ramsey, Rev. Mod. Phys. \textbf{62}, 541 (1990).

\bibitem{Haeffner2008} H. H{\"a}ffner, C. Roos, and R. Blatt, Phys. Rep. \textbf{469}, 155 (2008).

\bibitem{Haeffner2003} H. H{\"a}ffner, S. Gulde, M. Riebe, G. Lancaster, C. Becher, J. Eschner, F. Schmidt-Kaler, and R. Blatt, Phys. Rev. Lett. 90, 143602 (2003).

\bibitem{Millerioux1994} Y. Millerioux, D. Touahri, L. Hilico, A. Clairon, R. Felder, F. Biraben, and B. de Beauvoir, Opt. Commun. \textbf{108}, 91 (1994).

\bibitem{Badr2006} T. Badr, M. D. Plimmer, P. Juncar, M. E. Himbert, Y. Louyer, and D. J. E. Knight, Phys. Rev. A \textbf{74}, 062509 (2006).

\bibitem{Zanon2006} T. Zanon-Willette, A. D. Ludlow, S. Blatt, M. M. Boyd, E. Arimondo, and J. Ye, Phys. Rev. Lett. \textbf{97}, 233001 (2006).

\bibitem{Parthey2011} C. G. Parthey, A. Matveev, J. Alnis, B. Bernhardt, A. Beyer, R. Holzwarth, A. Maistrou, R. Pohl, K. Predehl, T. Udem, T. Wilken, N. Kolachevsky, M. Abgrall, D. Rovera, C. Salomon, P. Laurent, and T. W. H{\"a}nsch, Phys. Rev. Lett. \textbf{107}, 203001 (2011).

\bibitem{King2012} S. A. King, R. M. Godun, S. A. Webster, H. S. Margolis, L. A. M. Johnson, K. Szymaniec, P. E. G. Baird, and P. Gill, New J. Phys. \textbf{14}, 013045 (2012).

\bibitem{Huntemann2012} N. Huntemann, M. Okhapkin, B. Lipphardt, S. Weyers, Chr. Tamm, and E. Peik, Phys. Rev. Lett. \textbf{108}, 090801 (2012).

\bibitem{Taichenachev2006} A. V. Taichenachev, V. I. Yudin, C. W. Oates, C. W. Hoyt, Z. W. Barber, and L. Hollberg, Phys. Rev. Lett. \textbf{96}, 083001 (2006).

\bibitem{Baillard2007} X. Baillard, M. Fouch{\'{e}}, R. L. Targat, P. G. Westergaard, A. Lecallier, Y. L. Coq, G. D. Rovera, S. Bize, and P. Lemonde, Opt. Lett. \textbf{32}, 1812 (2007). 

\bibitem{Poli2008} N. Poli, Z. W. Barber, N. D. Lemke, C. W. Oates, L. S. Ma, J. E. Stalnaker, T. M. Fortier, S. A. Diddams, L. Hollberg, J. C. Bergquist, A. Brusch, S. Jefferts, T. Heavner, and T. Parker, Phys. Rev. A \textbf{77}, 050501 (2008). 

\bibitem{Akatsuka2010} T. Akatsuka, M. Takamoto, and H. Katori, Phys. Rev. A \textbf{81}, 023402 (2010).

\bibitem{Yudin2010} V. I. Yudin, A. V. Taichenachev, C. W. Oates, Z. W. Barber, N. D. Lemke, A. D. Ludlow, U. Sterr, C. Lisdat, and F. Riehle, Phys. Rev. A \textbf{82}, 011804(R) (2010). 

\bibitem{Hollberg1984} L. Hollberg and J. L. Hall, Phys. Rev. Lett. \textbf{53}, 230 (1984).

\bibitem{Taichenachev2009} A. Taichenachev, V. Yudin, C. Oates, Z. Barber, N. Lemke, A. Ludlow, U. Sterr, C. Lisdat, and F. Riehle, JETP Lett. \textbf{90}, 713 (2009).

\bibitem{Warren1983} W. S. Warren and A. H. Zewail, J. Chem. Phys. \textbf{78}, 2279 (1983).

\bibitem{Ramsey1951} N. F. Ramsey and H. B. Silsbee, Phys. Rev. \textbf{84}, 506 (1951).

\bibitem{Letchumanan2004} V. Letchumanan, P. Gill, E. Riis, and A. G. Sinclair, Phys. Rev. A \textbf{70}, 033419 (2004).

\bibitem{Tamm2009} Chr. Tamm, S. Weyers, B. Lipphardt, and E. Peik, Phys. Rev. A \textbf{80}, 043403 (2009).

\bibitem{Lipphardt2009} B. Lipphardt, G. Grosche, U. Sterr, Chr. Tamm, S. Weyers, and H. Schnatz, IEEE Trans. Instr. Meas. \textbf{58}, 1258 (2009).

\bibitem{Weyers2001} S. Weyers, U. H{\"u}bner, R. Schr{\"o}der, Chr. Tamm, and A. Bauch, Metrologia \textbf{38}, 343 (2001).

\end{thebibliography}
\end{document}